\documentclass[a4paper,oneside,11pt,english]{article}

\usepackage{braket,graphicx}
\usepackage{color}
\usepackage{authblk}

\topmargin  - 10mm

\title{\huge{Integrated AlGaAs source of highly indistinguishable and energy-time entangled photons}}

\author[1]{\vspace*{8mm}Claire Autebert}
\author[2]{Natalia Bruno}
\author[2]{Anthony Martin}
\author[3]{Aristide Lemaitre}
\author[3]{Carmen Gomez Carbonell}
\author[1]{Ivan Favero}
\author[1]{Giuseppe Leo}
\author[2]{Hugo Zbinden}
\author[1,*]{Sara Ducci\vspace*{5mm}}

\affil[1]{\emph{Laboratoire Mat\'eriaux et Ph\'enom\`enes Quantiques, Universit\'e Paris Diderot, Sorbonne Paris Cit\'e, CNRS-UMR 7162, Case courrier 7021, 75205 Paris Cedex 13, France}}
\affil[2]{\emph{Group of Applied Physics, University of Geneva, Switzerland}}
\affil[3]{\emph{Laboratoire de Photonique et Nanostructures, CNRS-UPR20, Route de Nozay, 91460 Marcoussis, France}}

\affil[*]{Corresponding author: sara.ducci@univ-paris-diderot.fr}

\date{}

\begin{document}

\maketitle

\begin{abstract}
The generation of nonclassical states of light in miniature chips is a crucial step towards practical implementations of future quantum technologies. Semiconductor materials are ideal to achieve extremely compact and massively parallel systems and several platforms are currently under development. In this context, spontaneous parametric down conversion in AlGaAs devices  combines the advantages of room temperature operation, possibility of electrical injection and emission in the telecom band. Here we report on a chip-based AlGaAs source, producing indistinguishable and energy-time entangled photons with a brightness of $7.2\times10^6$ pairs/s and a signal-to-noise ratio of  $141\pm12$. Indistinguishability between the photons is demonstrated via a Hong-Ou-Mandel experiment with a visibility of  $89\pm3\%$, while energy-time entanglement is tested via a Franson interferometer leading to a value for the Bell parameter $ S=2.70\pm0.10$. 
\end{abstract}

\newpage
Photonics is playing a key role in the development of future quantum technologies. Thanks to their propagation speed and long coherence time, photons are the most performing support for quantum communications~\cite{Gisin2007}. Moreover, integrated photonics technology enables the development of on-chip platforms for demanding applications of quantum simulation~\cite{Aspuru-Guzik2012}, computation~\cite{Crespi2011}  and  metrology~\cite{Giovannetti2011}, and allows to solve critical problems in terms of scalability and reliability~\cite{Tanzilli2012}. 
In this context, semiconductor materials may play a central role; in particular, the generation of entangled photon pairs has been demonstrated using biexciton cascade in quantum dots~\cite{ToshibaNJP2006,Senellart2010} and nonlinear interaction in AlGaAs~\cite{Orieux2013,JenneweinSciRep2013,Sarrafi14} or Silicon-based devices~\cite{Harada08,Takesue2014,Grassani15}.   
The target property of the generated quantum state depends on the intended application: photons indistinguishability is at the heart of controlled logic gates~\cite{Politi2009}, quantum networking~\cite{Kimble2008} and boson sampling~\cite{Spring2013a}. Entanglement is used to speed up algorithms~\cite{Knill2010}, protect encoded information~\cite{Ekert1991}, teleport quantum states~\cite{VanHouwelingen2006} and reduce intrinsic uncertainty in interferometry measurement~\cite{Humphreys2013}.
For these reasons, sources producing different quantum states while maintaining a high degree of compactness and integratabilty are highly desirable.
GaAs and its material derivatives like AlGaAs present a strong case to miniaturize different quantum components in the same chip. Its direct band-gap has already led to the monolithic integration of the primary laser source and the nonlinear medium into a single device emitting photon pairs under electrical injection at room temperature \cite{Boitier2014}. Moreover, GaAs strong electro-optical Pockels effect enables a fast control and manipulation of the generated photons as recently demonstrated~\cite{Wang2014}. On the front of on-chip single photon detection as well, high-efficiency superconducting nanowire single-photon detectors have been integrated with GaAs waveguides~\cite{Sprengers2011}. All these achievements consolidate the potential of this platform to realize miniature chips containing the generation, manipulation and detection of quantum states of light. 

In this letter, we present an AlGaAs ridge waveguide producing highly indistinguishable and entangled photon pairs at telecom wavelengths and room temperature. Our device has been optimized for efficient type-II spontaneous parametric down conversion (SPDC); two Bragg mirrors provide both a photonic band gap confinement for a TE Bragg mode at 783 nm~\cite{Yeh1976,Helmy06} and total internal reflection claddings for TE$_{00}$ and TM$_{00}$ modes at 1.56 $\mu$m. The sample is grown by molecular beam epitaxy on a (100) GaAs substrate. It consists of a 6-period Al$_{0.80}$Ga$_{0.20}$As/Al$_{0.25}$Ga$_{0.75}$As Bragg reflector (lower mirror), a 298\,nm Al$_{0.45}$Ga$_{0.55}$As core  and a 6-period Al$_{0.25}$Ga$_{0.75}$As/Al$_{0.80}$Ga$_{0.20}$As Bragg reflector (upper mirror).  Waveguides are fabricated using wet chemical etching to define 5.5-6\,$\mu$m wide and 5\,$\mu$m deep ridges along the (011) crystalline axis, in order to exploit the maximum non-zero optical nonlinear coefficient and a natural cleavage plane. Optical propagation losses in the telecom range of 0.3 and 0.5 \,cm$^{-1}$ for the TE$_{00}$ and TM$_{00}$ modes, respectively, are measured via a standard Fabry-Perot technique~\cite{DeRossi2005}.

\begin{figure}[h!]
	\centering
	\fbox{\includegraphics[width=10cm]{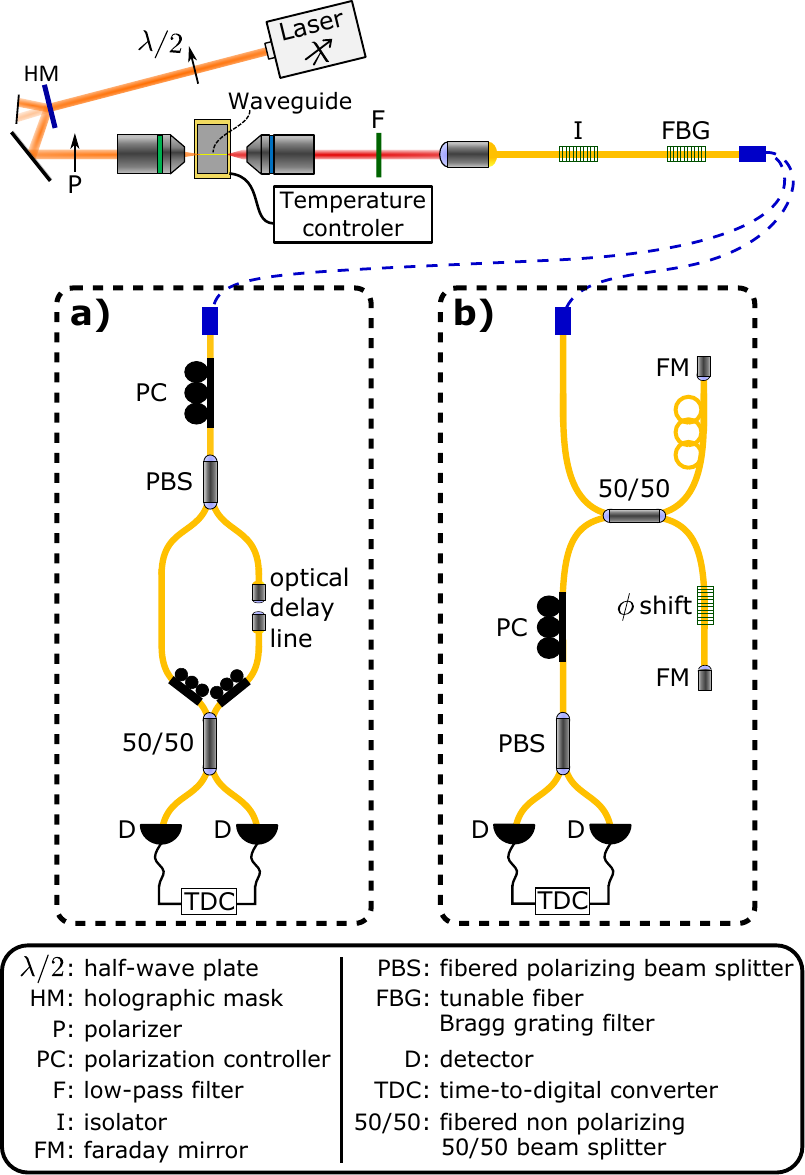}}
	\caption{Sketch of the experimental setup used for the HOM experiment (a) and for the Franson experiment (b). The pumping scheme and the collection of the photon pairs is common to the two experiments. For the HOM experiment the photon pairs are deterministically separated with a polarizing beam splitter and then recombined in a 50/50 beam splitter. An optical delay line allows to vary the relative arrival time of the two photons. For the Franson experiment the two entangled photons are directed into an unbalanced interferometer. A piezo actuator is used to control the relative phase $\phi$ between its short and long arm.
		}
	\label{Fig1}
\end{figure}

The ability of our device to produce indistinguishable photons is tested through a Hong-Ou-Mandel (HOM) experiment \cite{Hong1987}; in this type of measurement, two indistinguishable photons enter a 50/50 beam splitter at the same time. The destructive quantum interference makes them exit the beam splitter through the same output, thus inducing a dip in the coincidence histogram. Figure \ref{Fig1} shows a sketch of the experimental setup used for this experiment. The light beam of a cw Ti:Sa laser is used to excite the Bragg mode of the sample; after a spatial shaping with an holographic mask, it is injected into the waveguide with a microscope objective. Light emerging from the opposite end is collected with a second microscope objective, a fiber coupler and filtered with a tunable fibered Bragg grating (FBG) having a full width at half maximum (FWHM) of 10.8 nm.

\begin{figure}[htbp]
	\centering
	\fbox{\includegraphics[width=12cm]{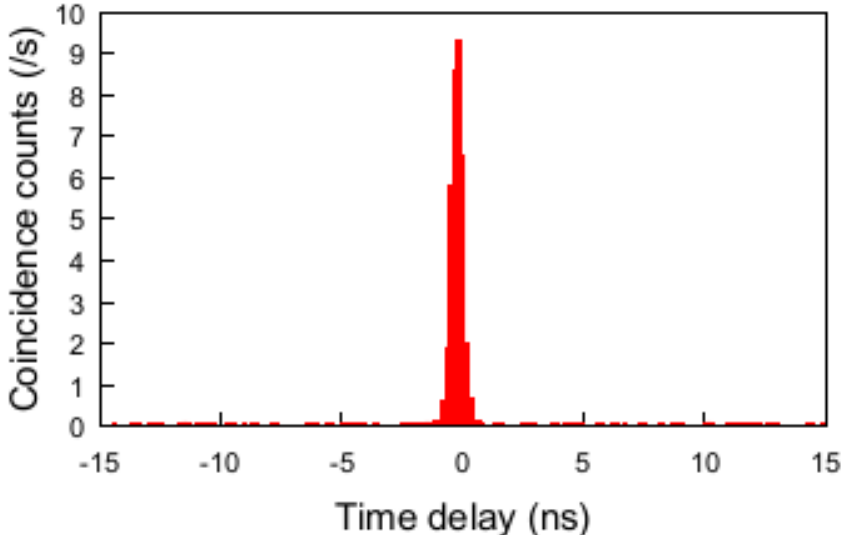}}
	\caption{Coincidence histogram of TE/TM photons passing through the FBG filter centered at 1.566\,$\mu$m at T$=20.1^{\circ}$C. The data were accumulated during 100\,s with a sampling resolution of 164\,ps. 
		}
	\label{Fig2}
\end{figure}

The optimum working point of our source is determined by measuring the temporal correlations between the TE and TM photons. Two stirling-cooled free running single photon avalanche photodiodes connected to a time-to-digital converter (TDC) are used for coincidence counting~\cite{Korzh2014}. Figure~\ref{Fig2} shows a histogram of the recorded detection time delays at temperature $T= 20.1^\circ$C for an internal estimated pump power of 625\,$\mu$W in the guided mode. 
The coincidence to accidental ratio (CAR), an important figure of merit for a photon pairs source, is calculated by taking the number of true coincidences within the FWHM of the peak over the background signal, on the same time window taken apart from the peak. 
The dependence of the CAR on both the pump wavelength and the internal pump power has been studied:  a maximum value of the CAR of $141\pm12$ is obtained for a pump wavelength around 783 nm and an internal pump power around 625\,$\mu$W, leading to a brightness of ${7.2\times10^6}$ pairs/s. This working point correspond to the phase matching resonance of the device: for a pump wavelength above the degeneracy point, the phase-matching condition is no longer satisfied and below degeneracy a shift of 0.2 nm results in a variation of 100\,nm for the signal and idler wavelengths, which are by consequence outside the interferential filter. The CAR value, limited by the detectors dark counts, is the maximum ever obtained on a semiconductor waveguide up to our knowledge; this is due to the low value of optical losses of our sample and the low level of noise of the detectors employed in this work.

\begin{figure}[h!]
	\centering
	\fbox{\includegraphics[width=12cm]{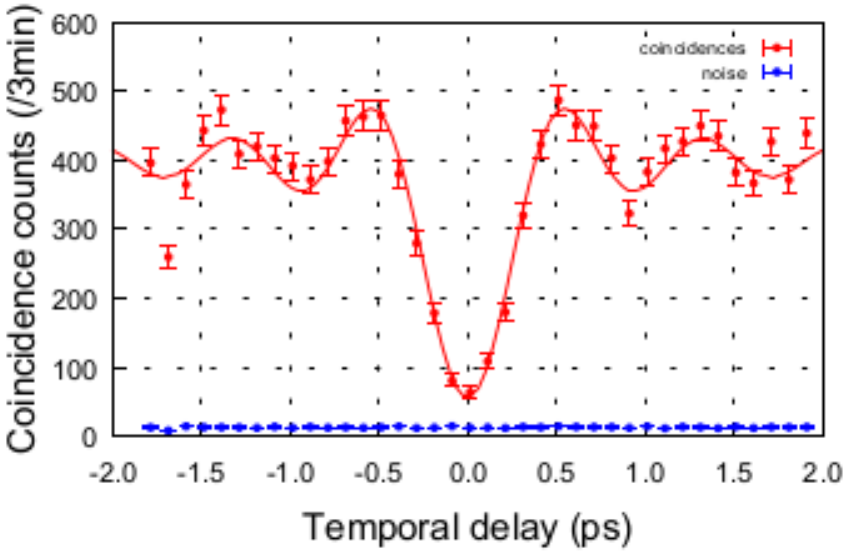}}
	\caption{Results of the HOM-type experiment. The uncertainty associated with each point has been calculated using standard squared root deviation. The line shows the adjustment of the data with equation \ref{equation1}. The obtained net visibility is $V_{net}=89.0\pm2.8\%$.}
	\label{Fig3}
\end{figure}

After the identification of the optimum working point, we proceeded to the HOM measurement; for this we used a polarization controller to align the polarizations of the photons and we inserted an interferometer with an optical delay line on one of the two arms, followed by a a 50/50 beam splitter before the detectors.
Figure~\ref{Fig3} reports the dip observed in the coincidence counts as a function of the optical path length difference between the two arms of the interferometer: this dip is a clear signature of the destructive quantum interference between the two photons. The shape of the dip is given by the convolution of the two wavepackets arriving at the 50/50 beam splitter. Since the signal/idler photons are filtered with a rectangular FBG filter we adjust the experimental data with the theoretical expression of the HOM dip: 
\begin{equation}
	N_c={\rm A}\left( 1- {\rm V}\times{\rm sinc} \left(2\pi\delta t\frac{ \delta\lambda c}{\lambda^2}\right)\right)
\label{equation1}
\end{equation}
where $N_c$ is the coincidence rate, ${\rm A}$ the coincidence rate apart from the bunching region, ${\rm V}$ the visibility,  $\delta t$ the time delay and $\delta \lambda$ the FWHM spectral intensity. 
The two fitting parameters are  $\delta\lambda$ and ${\rm V}$; for the first one we obtain $10.7\pm 0,2$ nm, in very good agreement with the FBG filter FWHM, while for the second one we obtain a net (raw) visibility of $89.0\pm 2.8\%$ ($86.1 \pm 2.7\%$). 
This result is an unprecedented value in a semiconductor waveguide; the limitation to the visibility in our experiment can be attributed to the reflectivity  R of our waveguide facets, which is around $24\%$ for the TE and TM modes.
Thus, a coincidence event can not only be due to two photons directly transmitted by the facets, but also to one photon directly transmitted and one photon having experienced two reflections before leaving the waveguide. Since in the latter case, the path difference for the two photons is not the same as for the former case, these photons do not contribute to the dip ~\cite{Caillet2010}.
In this case the maximum achievable visibility is  given by:
 \begin{equation}
		V = \frac{1}{1+\frac{R^2}{1-R} (\eta_{TM} + \eta_{TE})} = 90.5\%
\label{equation2}
\end{equation}
where $\eta_{TM}$ and $ \eta_{TE}$ correspond to the transmission efficiency for the two polarisation modes in the sample. This expression is in excellent agreement with our experimental results.
Standard telecom anti-reflection coating on the facets of AlGaAs waveguides allows to reach transmittivities of almost $100\%$; this kind of treatment applied to our device would thus increase the indistiguishability of the emitted photons. 

Among different possible entangled states, we have chosen to produce energy-time entangled photons; this is a very convenient format of entanglement, as it can be easily manipulated with  integrated circuits and can be preserved over long distances in standard optical fibers~\cite{Korzh2015}. By pumping the device with a continuous-wave laser, the photon pairs are emitted simultaneously, but their emission time is undetermined within the coherence time of the pump laser. This lack of information leads to energy-time entangled pairs, as first pointed out by Franson \cite{Franson1989}. We have thus implemented the experimental setup sketched in Figure~\ref{Fig1}(b). Before passing through the polarizing beam splitter, the two entangled photons are directed into an unbalanced interferometer. A piezo actuator is used to control the relative phase $\phi$ between its short and long arm. An essential condition to fulfill in a Franson type experiment is that $(\tau_c,\tau_{\rm det})<<\Delta t << \tau_p$, where $\tau_c$ is the coherence time of the signal and idler photons, $\tau_{\rm det}$ the jitter of the detectors, $\Delta t$ = $\Delta$L/c is the time difference between the two paths of the interferometer, and $\tau_p$ is the pump laser coherence time. As shown in Figure~\ref{Fig3}, the coherence time of the photons is 0.7 ps. We have thus chosen a path-length difference of the interferometer of 2.5 ns, which is also much smaller than the 1$\mu$s of coherence time of the cw laser pump (TOPTICA DL 100) and bigger than the timing jitter of the detectors (200 ps).

\begin{figure}[htbp]
	\centering
	\fbox{\includegraphics[width=12cm]{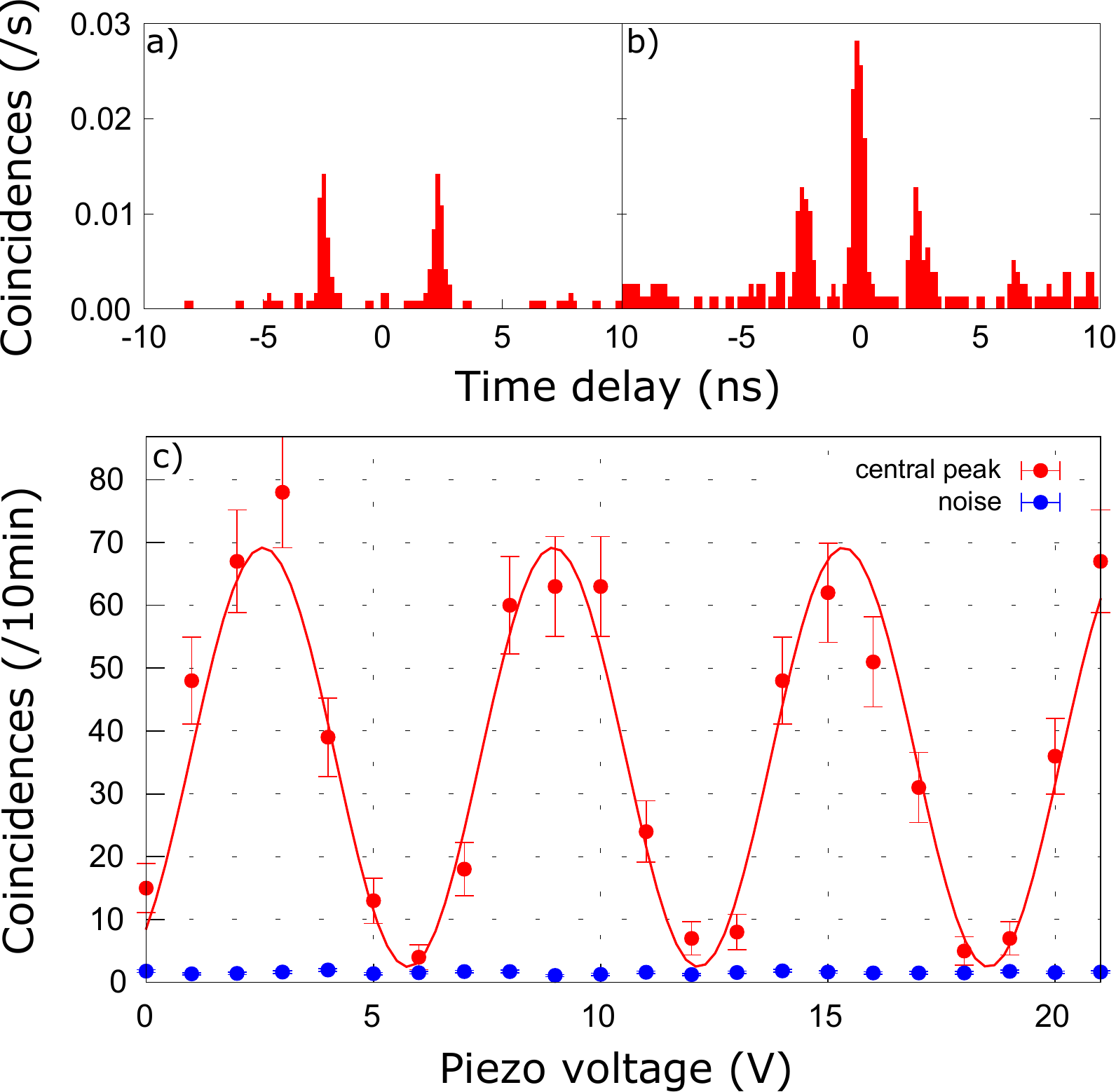}}
	\caption{Results of the Franson experiment: 
		a) and b) histograms of the coincidence rate for two different phase settings, chosen to minimize and maximize the central peak ;
		c) two photon interference: the coincidence count rate of the central peak is plotted as a function of the phase $\phi$. The internal estimated pump power is 273\,$\mu$W. The line show the fitting curve used to estimate the visibility $V_{net}=95.6\pm3.7\%$.
		}
	\label{Fig4}
\end{figure}

As shown in Figure~\ref{Fig4}(a) and (b) the recorded histogram has three coincidence peaks; the left and right peaks correspond to a situation when one photon goes through the short arm ($\rm s$) and the other through the long arm ($\rm l$) of the interferometer. The middle peak results from the interference between the state where both photons pass through the short arm and the one where both photons pass through the long arm. This peak results from the quantum interference of the following post selected state : 
\begin{equation}
	\ket{\Phi}= \frac{1}{\sqrt{ 2}}[\ket{ \rm s}_i\ket{\rm s}_s + e^{-i2\phi}\ket{\rm l}_i\ket{\rm l}_s]
\label{equation3}
\end{equation}
where the indices $s$ and $i$ stand for signal and idler.

For perfectly entangled photons the total number of coincidences in the central peak exhibits an interference fringe of unity visibility when the phase $\phi$ is varied. The two values of $\phi$ for the measurements reported in Figure~\ref{Fig4}(a) and (b) have been chosen to minimize and maximize the central peak height, respectively. The height of the satellite peaks is independent of the phase. The measurement accuracy is 200\,ps, dominated by the jitter of the detectors.
The coincidence counts corresponding to the central peak as a function of $\phi$ are plotted in Figure~\ref{Fig4}(c). The trend is well fitted by a sinusoidal curve having a net (raw) visibility of  $95.6\pm 3.7\%$ ($91.5 \pm 3.6\%$). For the net (raw) data, this leads to a value of the Bell parameter S of $2.70 \pm 0.10$ ($2.58 \pm 0.10$) and a consequent violation of Bell's inequality by 6.7 (5.8) standard deviations. 

In conclusion, we have demonstrated an AlGaAs device working at room temperature with proven compliancy with electrical pumping \cite{Boitier2014}, generating photon pairs linked by a high degree of entanglement and able to interfere on a beam-splitter with high visibility. The only solid-state source featuring both the last two properties demonstrated up to now is based on single quantum dot technology ~\cite{Muller2013}; with respect to that result our device present the advantage of working at room temperature and displaying higher values of HOM and Bell visibilities. The present results constitute a step towards on-chip large scale photonic circuit-based quantum computation and simulation. They set the AlGaAs platform in an advantageous position for the development of monolithic complex architectures exploiting photon pairs generation by electrical injection and fast quantum state manipulation via electro-optics effect. Moreover, the recent progress on hybrid integration of III-V compound semiconductors onto silicon-on-insulator substrates allows combining the advantages of our device to a nearly complete suite of silicon photonics components. These include filters, (de)multiplexers, splitters, interferometers and photodetectors for the fabrication of novel generations of CMOS compatible chips for quantum information technologies.

\section*{Funding Information}
Agence Nationale de la Recherche (ANR-14-CE26-0029-01); Institut Universitaire de France; D\'el\'egation G\'en\'erale de l'Armement; R\'egion Ile-de-France in the framework of DIM Nano-K; Swiss National Science Fondation NCCR-QSIT; French RENATECH.

\section*{Acknowledgments}
The authors thank B. Korzh for help with the stirling-cooled free running single photon avalanche photodiodes and Carlos Eduardo R. de Souza for the fabrication of the holographic mask.


\end{document}